\documentclass[11pt]{article}

\usepackage[preprint]{acl}

\usepackage{times}
\usepackage{latexsym}
\usepackage[T1]{fontenc}
\usepackage{amsmath}
\usepackage{amsfonts}
\usepackage{stmaryrd}
\usepackage{mathtools}
\usepackage[ruled,vlined,linesnumbered]{algorithm2e}
\usepackage[dvipsnames]{xcolor}
\usepackage{enumitem}
\usepackage{adjustbox}
\usepackage{caption}
\usepackage{booktabs}
\usepackage{subcaption}
\usepackage{multirow}
\usepackage{pgfplots}
\usepackage{hyperref}
\usepgfplotslibrary{fillbetween}
\usepgfplotslibrary{external}
\usepgfplotslibrary{groupplots}
\usetikzlibrary{calc}

\pgfplotsset{compat=1.18}

\usepackage[utf8]{inputenc}
\usepackage{microtype}
\usepackage{inconsolata}
\usepackage{graphicx}

\usepackage{pifont}
\newcommand{\xmark}{\ding{55}}%

\usepackage{amsmath}
\usepackage{amsthm}

\newtheorem{proposition}{Proposition}

\title{FOSTER: First-order Dataset Distillation for Text-based Sequential Recommendation}

\author{
 \textbf{Hung Vinh Tran\textsuperscript{1}},
 \textbf{Tong Chen\textsuperscript{1}},
 \textbf{Xinyi Gao\textsuperscript{1}},
 \textbf{Junliang Yu\textsuperscript{2}},
 \textbf{Julien Monteil\textsuperscript{3}},
 \textbf{Hongzhi Yin\textsuperscript{1}},\\
 \textsuperscript{1}The University of Queensland,
 \textsuperscript{2}Griffith University,
 \textsuperscript{3}Amazon International Machine Learning,
\\
 \small{
   \{h.v.tran,tong.chen,xinyi.gao,h.yin1\}@uq.edu.au; junliang.yu@griffith.edu.au; jul@amazon.com
 }
}

\begin{document}
\maketitle
\begin{abstract}

Text-based sequential recommender systems, while greatly improving recommendation accuracy by incorporating item contexts, are undeniably more expensive to train. 
By condensing a large dataset into a compact set of synthetic samples for model training, dataset distillation offers a promising solution. 
However, its adoption in text-based sequential recommendation is non-trivial given the large pool of discrete items. This challenge is further compounded by language model-based item encoding, which makes bi-level optimization commonly used in dataset distillation prohibitively expensive.
To this end, we propose \textbf{F}irst-\textbf{o}rder data\textbf{s}et distillation for \textbf{Te}xt-based Sequential \textbf{R}ecommendation (\textbf{FOSTER}), which facilitates effectiveness and efficiency via three novel components: 
(1) stochastic item subset sampling that replaces costly full-corpus embedding extraction at each distillation step;
(2) first-order optimization with trajectory-anchored parameter reset to avoid expensive bi-level gradient computation; 
and 
(3) regularization that explicitly promotes co-occurrence between semantically similar items in the synthetic sequences.
Extensive experiments on three benchmarks show that FOSTER consistently outperforms existing dataset distillation and coreset selection baselines, approximating full-dataset performance 
using as few as 20 synthetic interaction sequences.
\end{abstract}

\section{Introduction}


Sequential recommendation is the task of suggesting new, unvisited items based on a user's historical interaction sequence. Unlike traditional ID-based methods that directly map item IDs into embeddings for modeling interaction sequences, text-based sequential recommenders leverage language models such as BERT~\cite{morec2023} and LLMs~\cite{liu2024once} to encode item textual metadata (e.g., titles and descriptions). With such semantic-aware representations, text-based sequential recommenders benefit from improved generalizability and cold-start robustness, positioning them as a strong direction for next-generation recommender systems~\citep{morec2023}.

However, these benefits often come with trade-offs in training efficiency, which is vital for recommendation models as they require periodic updates~\cite{lee2023important}. 
This is due to not only the increased data size as each item is now represented by a set of tokens instead of a single ID, but also the overhead from the additional text encoder -- commonly a heavy language model -- for generating item embeddings.
A natural solution is to reduce the training data volume, where existing methods fall into two main categories.
Coreset selection methods identify a small, representative subset of the original dataset, preserving real interactions that best capture the underlying data distribution \citep{dealrec2024, mei2025goracs, tran2026nacs}.
Dataset distillation, by contrast, generates a compact set of synthetic training samples that compress information learned from the entire dataset, resulting in a more information-dense dataset and superior model performance.
These benefits have motivated recent efforts to extend dataset distillation to recommender systems, such as CGM~\cite{wang2023gradient} and TD3~\cite{zhang2025td3}.

Despite the promises, applying dataset distillation to text-based recommendation introduces significant challenges. 
First, existing methods like TD3 \cite{zhang2025td3} represent each synthetic interaction as a probability over the entire item space, instead of discrete items in real sequences. 
Though this ensures coverage of all items, replacing discrete items with dense distributions 
requires computing all item embeddings for every synthetic interaction. The computational cost is further amplified in text-based recommendation, where each item is represented by a collection of tokens instead of a single ID. This gives rise to the first challenge:  \textbf{(C1) How to distill quality synthetic sequences from the large pool of items and text tokens?}

Second, dataset distillation is typically formulated as a bi-level optimization problem \citep{wang2018dataset,zhang2025td3} that iteratively refines the synthetic data conditioned on a downstream model trained in the inner loop,  therefore incurring high computational costs \citep{yu2023dataset}. 
Though recent advances in vision tasks introduce first-order alternatives~\citep{yin2023squeeze,zhao2023dataset}, these methods are constrained to a small and fixed label space (e.g., object classes) when optimizing synthetic samples. 
This assumption breaks down in recommendation, where the prediction space corresponds to a large item catalog and each item may serve as a distinct target. Since existing first-order methods optimize separate synthetic samples for each class, their synthetic set size and computational cost scale linearly with the number of items, making them prohibitively expensive and unsuitable for recommendation.
Therefore, the second challenge is: \textbf{(C2) Can dataset distillation escape the heavy computation trap of bi-level optimization for text-based sequential recommendation?}

Third, recommender models commonly use tied embeddings, i.e., sharing the same item embedding space for input encoding and output prediction. This practice is theoretically justified only when the underlying data satisfies the distributional hypothesis, where semantically related items tend to appear in similar contexts~\citep{pmlr-v235-bertolotti24a}. However, existing recommendation dataset distillation methods overlook the associative patterns among co-occurring items when compress interactions, leading to loss of important semantics that are crucial to recommendation tasks. Thus, the third challenge is:
\textbf{(C3) How to preserve items' co-occurrence semantics in synthetic sequences to fully capitalize embedding tying?}

\begin{figure*}
    \centering
    \subcaptionbox{Pipeline Overview\label{fig:overview}}{%
    \includegraphics[width=0.58\linewidth]{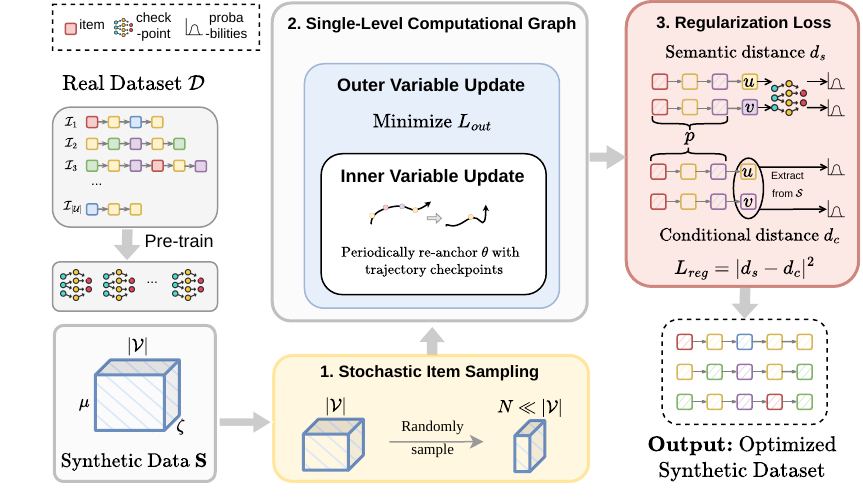}%
    }
    \hfill
    \subcaptionbox{BOME compared to BPTT\label{fig:bomevsbptt}}{%
    \includegraphics[width=0.4\linewidth]{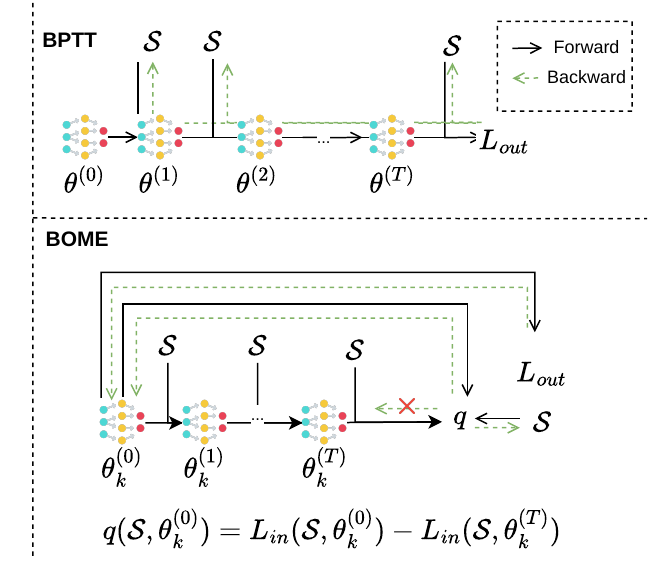}%
    }
    \vspace{-0.2cm}
    \caption{(a) Overview of our proposed pipeline. Given real dataset $\mathcal{D}$
, we distill it into a compact synthetic dataset $\mathcal{S}$ via stochastic item sampling, a single-level computational graph with inner/outer variable updates, and a regularization loss $L_{reg}$. (b) Comparison of BOME and BPTT unrolling strategies, illustrating how BOME avoids full trajectory backpropagation by using constraint $q$.}
    \label{fig:placeholder}
\end{figure*}

Motivated by these challenges, we propose 
\underline{f}irst-\underline{o}rder data\underline{s}et distillation for \underline{te}xt-based sequential \underline{r}ecommendation (\textbf{FOSTER}).
Specifically, for \textbf{C1}, rather than involving all items at each training step, we instead sample a small random subset of items per batch and compute sequence probabilities directly from their embeddings via Tucker decomposition \cite{tucker1964extension}, significantly reducing the per-step computational cost. 
This is similar with negative sampling: rather than optimizing against the full item catalog at each step as in full softmax, we sample a random subset of items per step.
To tackle \textbf{C2}, inspired by BOME \cite{liu2022bome}, we propose a first-order optimization strategy for sequential recommendation, which jointly optimizes the inner (model weight) and outer (synthetic dataset) parameters. 
To keep the inner model well-behaved, we employ trajectory-anchored regularization that periodically reverts them to checkpoints sampled from the real training trajectory, preventing drift toward degenerate states unreachable by any realistic training run.
Lastly, for \textbf{C3}, we introduce a regularization term that encourages the synthetic dataset to satisfy the distributional hypothesis, enabling more reliable dataset results. Figure \ref{fig:overview} shows the overview of our method and our contributions are as follows:
\begin{enumerate}[noitemsep, topsep=0pt]
    \item We formalize the challenges of dataset distillation for text-based sequential recommendation, highlighting the key challenges: (1) expensive full corpus calculation at each time-step, (2) the bi-level trap for text-based sequential recommendation, and (3) the misalignment in preserving the original items' co-occurrence distribution in synthetic datasets.
    \item We propose \textsc{FOSTER}, an efficient dataset distillation framework for text-based sequential recommendation. It improves scalability through stochastic item subset sampling, stabilizes first-order bi-level optimization with trajectory-anchored regularization, and preserves item co-occurrence semantics under embedding tying.
    \item Extensive experiments on $3$ representative datasets, namely: Games, Foods \cite{ni2019justifying}, and Yelp, show that we can approximate the model performance using only 20 synthetic sequences for Games and Foods, and 60 sequences for Yelp. Furthermore, FOSTER is significantly more efficient and effective than existing recommendation dataset distillation methods. 
\end{enumerate}

\section{Preliminaries}
In this section, we will provide preliminary on text-based sequential recommendation and dataset distillation. After that, we provide the initial pipeline for sequential recommendation introduced by \citet{zhang2025td3}.

\subsection{Text-based Sequential Recommendation}

Given user set $\mathcal{U}$ and item set $\mathcal{V}$, with each item $v_i$ is associated with a textual description $d_i$, such as title or item description, from which a dense embedding vector $e_i$ is extracted through a language model $l(\cdot, \theta)$, i.e. $\mathbf{e}_i=l(d_i, \theta)$.
For each user $u$, let $\mathcal{I}_u=\left[ v_1, v_2, ..., v_n \right]$ denotes their chronological ordered interaction sequences. 
A sequential recommendation model (SeqRec) $g_\phi$ consumes the sequence of item embeddings $[\mathbf{e}_{v_1}, \ldots, \mathbf{e}_{v_{|\mathcal{I}_u|}}]$ and produces a user representation $\mathbf{h}_u = g_\phi(\mathbf{e}_{v_1}, \ldots, \mathbf{e}_{v_{|\mathcal{I}_u|}})$. 
At inference, the relevance score between user $u$ and a candidate item $v$ is computed as the dot product $s(u, v) = \mathbf{h}_u^\top \mathbf{e}_v$, and the top-$K$ items are recommended accordingly. Both $l_\theta$ and $g_\phi$ are trained end-to-end by minimizing a binary cross-entropy (BCE) loss over observed and sampled negative interactions, creating the final recommendation model $f=l \cdot g$.

\subsection{Dataset Distillation}
Given a large training dataset $\mathcal{D} = \{\mathcal{I}_u\}_{u \in \mathcal{U}}$, dataset distillation \citep{yu2023dataset,zhang2025td3} aims to synthesize a compact synthetic dataset $\mathcal{S}$  consisting of $\mu$  sequences, each of maximum length $\zeta$, such that $|\mathcal{S}| \ll |\mathcal{D}|$, and a model trained on $\mathcal{S}$ achieves performance comparable to one trained on $\mathcal{D}$. 
Formally, this is defined as a bi-level optimization problem:
\begin{equation}
\label{eq:ori-opt-obj}
\begin{aligned}
            &\mathcal{S}^* = \arg\min_{\mathcal{S}} L_{\text{out}} \left(\theta^*(\mathcal{S}),\, \mathcal{D} \right) \\ 
        &\text{ s.t. } \theta^*(\mathcal{S}) = \arg\min_{\theta}\, L_{\text{in}}\!\left(\theta,\, \mathcal{S}\right),
\end{aligned}
\end{equation}
where $L_{\text{in}}$ is the recommendation loss function of the inner problem that learns SeqRec parameters $\theta$ on the synthetic data $\mathcal{S}$. $L_{\text{out}}$ is the outer loss function that evaluates $\theta$ on the real data $\mathcal{D}$, so as to facilitate the optimization of  $\mathcal{S}$. 

Since the exact solution $\theta^*(\mathcal{S})$ is intractable, Backpropagation Through Time (BPTT) \cite{wang2018dataset} approximates it by unrolling $T$ steps of gradient descent on the inner objective, and differentiating the outer loss through the entire unrolled computation graph. 
However, full unrolling with a fixed window suffers from short-horizon bias and unstable gradients. Random Truncated BPTT (RaT-BPTT)~\cite{feng2024embarrassingly} mitigates this by sampling a random window to train the inner loop, and truncated the last $T$ step only. However, it is still subsumed under a bi-level optimization formulation, inheriting the high computational cost.

\subsection{Sequential Recommendation Dataset Distillation}
Following~\cite{zhang2025td3}, we represent the synthetic dataset as a three-dimensional probabilistic tensor $\mathbf{S} \in \mathbb{R}^{\mu \times \zeta \times |\mathcal{V}|}$, where $\mathbf{S}_{ij:} \in \mathbb{R}^{|\mathcal{V}|}$ encodes a soft distribution over the entire item set for synthetic user $i$ at timestep $j$, allowing gradient-based optimization over the otherwise discrete interaction space.


Directly optimizing $\mathbf{S}$ is intractable when $|\mathcal{V}|$ is large, as the tensor size scales with $\mu \times \zeta \times |\mathcal{V}|$. To address this, $\mathbf{S}$ is parameterized via Tucker decomposition \cite{tucker1964extension}:
\begin{equation}
\label{eq:tucker}
    \mathbf{S} = \mathbf{G} \times_1 \mathbf{U} \times_2 \mathbf{T} \times_3 \mathbf{E} =: \llbracket \mathbf{G}; \mathbf{U}, \mathbf{T}, \mathbf{E} \rrbracket,
\end{equation}
where $\times_n$ indicates the tensor product along the n-th mode, $\mathbf{U} \in \mathbb{R}^{\mu \times d_1}$ is the synthetic user latent factor, $\mathbf{T} \in \mathbb{R}^{\zeta \times d_2}$ is the temporal dynamics latent factor, $\mathbf{E} \in \mathbb{R}^{|\mathcal{V}| \times d_3}$ is the item factor shared with the pretrained item embedding table (frozen during distillation), and $\mathbf{G} \in \mathbb{R}^{d_1 \times d_2 \times d_3}$ is the relation core with $d_1, d_2, d_3 \ll |\mathcal{V}|$.

In \cite{zhang2025td3}, the learnable parameters are optimized with bi-level optimization, where the detailed training objective is discussed in Appendix \ref{sec:optimization-seqrec}. Although the number of learnable parameters reduces drastically, as we mentioned above, this approach requires to compute all item probability in all inner steps, limiting its scalability for text-based sequential recommendation with large item and token sets. In what follows, we present our solution FOSTER that addresses this efficiency challenge while preserving data distillation quality. 

\section{Methodology}
As shown in Fig. \ref{fig:overview}, FOSTER consists of three novel components. First, the stochastic item sampling significantly reduces memory usage and computational time at each iteration. Second, a first-order algorithm is designed to bypass the high computational overhead of bi-level optimization. Third, a regularization loss is in place to explicitly enforce distributional hypothesis, preserving items' co-occurrence semantics and uplifting the synthetic data utility. We discuss their details below.

\subsection{Item Sampling}

A critical scalability bottleneck in Tucker-based distillation in Eq. \ref{eq:tucker} arises from the item dimension of $\mathbf{S}$, whose size grows linearly with $\vert \mathcal{V} \vert$. In the inner loop,  all item embeddings must be computed at every step, making distillation on large-catalog datasets prohibitively expensive. 
To address this, we propose stochastic item subset sampling, which approximates the full-item objective by uniformly drawing a subset of $N \ll  \vert \mathcal{V} \vert$  items at each step, which we empirically show that suffices to maintain distillation quality while reducing per-step complexity drastically.
Formally, at each inner step, we sample a subset $\mathcal{V}_k \subset \mathcal{V}$ of $N$ items and construct a sampled synthetic batch via:
\begin{equation}
    \mathbf{S}_k = \llbracket \mathbf{G}; \mathbf{U}, \mathbf{T}, \mathbf{E}_k \rrbracket,
\end{equation}
where $\mathbf{E}_k \in \mathbb{R}^{N \times d}$ is the reduced version of the pretrained item embedding table $\mathbf{E}$ of $\mathcal{V}_k$.
Crucially, we apply this approximation consistently across both optimization loops.

\subsection{First-order Bi-level Optimization for Dataset Distillation}
\label{sec:bome}

Although the objective in Eq.~\ref{eq:ori-opt-obj} has wide applicability, it is
a bi-level optimization problem and typically requires differentiating through
the inner training trajectory, leading to expensive 
second-order
gradients. Inspired by~\citep{liu2022bome}, we reformulate the bi-level
problem as a single-level constrained optimization problem for SeqRec.

\begin{proposition}[First-order Optimization]
\label{prop:bome}
Suppose $L_{\mathrm{in}}(\mathcal{S}, \theta)$ is continuously differentiable in both arguments, and 
$\theta^*(\mathcal{S}) \in \arg\min_{\theta} L_{\mathrm{in}}(\mathcal{S}, \theta)$. We define the constraint
\begin{equation}
q(\mathcal{S}, \theta) \coloneqq
L_{\mathrm{in}}(\mathcal{S}, \theta)
- L_{\mathrm{in}}(\mathcal{S}, \theta^*(\mathcal{S})).
\end{equation}
Then $q(\mathcal{S}, \theta) \geq 0$ with equality iff
$\theta \in \arg\min_{\theta} L_{\mathrm{in}}(\mathcal{S}, \theta)$.
The bi-level problem in Eq.~\ref{eq:ori-opt-obj} is equivalent to
\begin{equation}
\label{eq:bome-obj}
\min_{\mathcal{S}, \theta}\;
L_{\mathrm{out}}(\theta, \mathcal{S}, \mathcal{D})
\quad \text{s.t.} \quad q(\mathcal{S}, \theta) \leq 0,
\end{equation}
which can be solved by first-order gradients alone.
\end{proposition}

The proof is provided in Appendix~\ref{app:first_order}. In practice, the exact
$\theta^*(\mathcal{S})$ is unavailable. We approximate it by $\theta^T$, obtained after
$T$ gradient steps on the inner objective, and use
\begin{equation}
    \hat q(\mathcal{S},\theta)
    =
    L_{\mathrm{in}}(\mathcal{S},\theta)
    -
    L_{\mathrm{in}}(\mathcal{S},\mathrm{sg}(\theta^T))
    \label{eq:q_hat}
\end{equation}
to approximate $q(\mathcal{S},\theta)$, where $\mathrm{sg}(\cdot)$ denotes stop-gradient.
This prevents back-propagation through the construction of $\theta^T$ and
removes the expensive second-order computation.

To solve Eq.~\ref{eq:bome-obj}, we employ a dynamic barrier update. At the
$k$-th outer step, let $L_k=L_{\mathrm{out}}(\theta_k,S_k,D_k)$ and
$\hat q_k=\hat q(S_k,\theta_k)$. The update direction is
\begin{equation}
\label{eq:lambdk}
\begin{aligned}
    \delta_k &= \nabla L_k + \lambda_k \nabla \hat{q}_k, \\
    \phi_k &= \eta \|\nabla \hat{q}(\mathcal{S}_k, \theta_k)\|^2, \\
    \lambda_k &= \max\left(\frac{\phi_k - \langle \nabla L_k, \nabla \hat{q}_k \rangle}{\|\nabla \hat{q}_k\|^2},\ 0\right),  
\end{aligned}
\end{equation}
where $\eta=0.5$ controls the barrier strength. We iteratively update $\left( \mathcal{S}, \theta \right)$ to reduce $L_{out}$, while decreasing $q$ to ensure the constraint was held. 

\begin{figure}
    \centering
    \subcaptionbox{Entropy}{\includegraphics[width=0.48\linewidth]{
        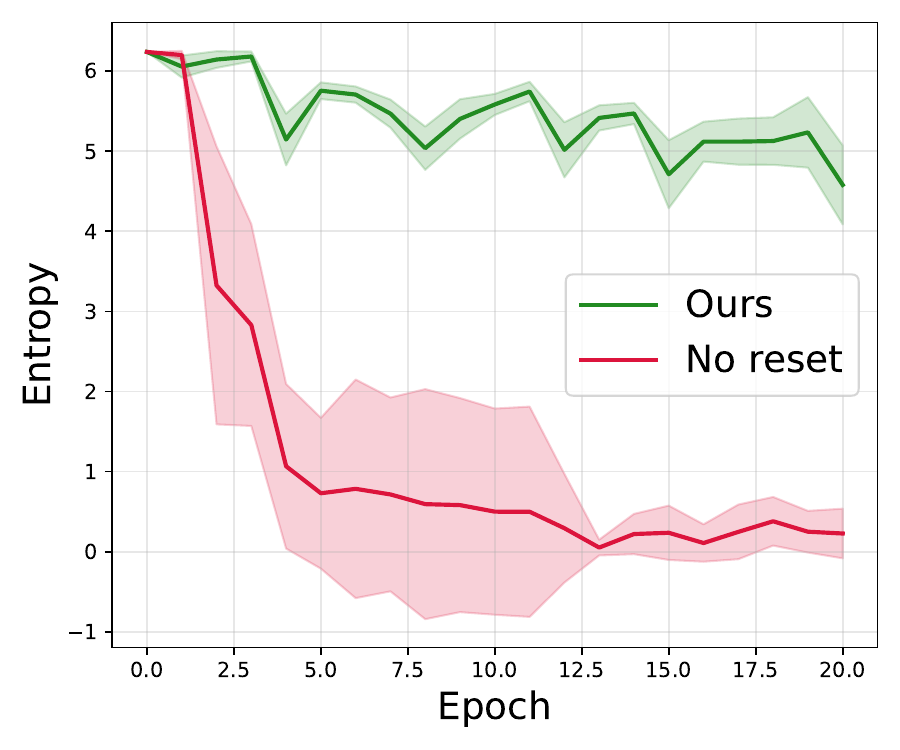
    }}
    \subcaptionbox{Training Trajectory}{\includegraphics[width=0.48\linewidth]{
        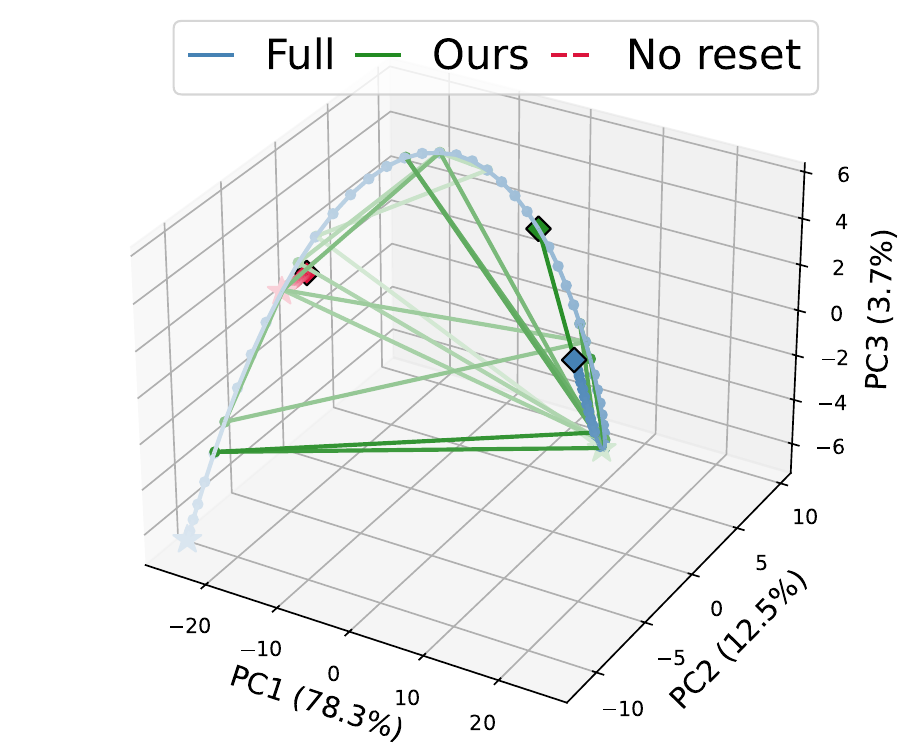
    }}
    \vspace{-0.2cm}
    \caption{Effect of periodic reset on bilevel distillation. (a) Average sequence entropy over 512 sampled items in distilled sequences. (b) Inner parameters $\theta$ in PCA decomposition of the full-data trajectory, our approach and using naive no reset approach. }
    \label{fig:reset-qualitative}
\end{figure}

However, naively applying such update to dataset distillation leads to degenerate solutions. Since the synthetic dataset $\mathcal{S}$ and the model parameters $\theta$
 are optimized jointly from scratch, $\mathcal{S}$ tends to overfit to the specific trajectory of $\theta$, resulting in a distilled dataset that fails to generalize to independently trained models. 
 To demonstrate this, we calculate the average entropy of the distilled sequences with a fixed 512 synthetic items in Games dataset over the first 20 training epoch. 
 As shown in Fig. \ref{fig:reset-qualitative}, $\mathcal{S}$ quickly collapses onto a narrow subset of items, while $\theta$ gradually drifts away from the original parameter space. 

To address this issue, we avoid maintaining a single $\theta$ during optimization. Instead, we anchor the inner variable by sampling from a pre-recorded training
trajectory. Let
$\mathcal{T}=\{\bar\theta_1,\bar\theta_2,\ldots,\bar\theta_M\}$ denote the anchor pool. After $R$ outer loops, we perform a reset step by sampling
$\bar\theta_k\sim\mathcal{T}$ and setting
$\theta_k\leftarrow\bar\theta_k$. 

 Our approach differs from MTT~\cite{cazenavette2022dataset}, which aim to match the original training trajectory, our method aim to improve the performance of the sampled checkpoint.
 And unlike naive joint optimization, $\theta_k$ is reset after a few outer iterations rather than inherited from the previous step. 
This constrains optimization to
valid and diverse model states observed during real training, decouples the
distilled data from a single co-evolving model, and encourages $\mathcal{S}$ to capture
transferable recommendation patterns.

\subsection{Regularization Loss for Distributional Hypothesis Alignment}
Recommendation models commonly adopt embedding tying, where the input and output sides share the same item embedding matrix. However, as shown by \citet{pmlr-v235-bertolotti24a}, this design is theoretically justified only when the distributional hypothesis (DH) holds. 
However, there is no guarantee that the distillation process naturally produces sequences that satisfy this property. 
Thus, synthetic data may distort item-level conditional distributions, making semantically similar items conditionally inconsistent. 
This causes DH to fail, so the tied embedding matrix receives conflicting optimization signals and leads to unstable training and degraded accuracy.

To address this, we introduce a regularization loss $L_{r}$
 that explicitly encourages the synthetic dataset to satisfy the distributional hypothesis. The core idea is to align two notions of item similarity -- \textbf{semantic equivalent} and \textbf{conditional equivalent} -- for any randomly sampled pair of items $(u, v)$. 
 
 The first is the semantic distance $d_s(u, v)$, where we use the fixed pretrained model $f$ as a proxy to judge whether $u$ and $v$ are semantically similar: if substituting one for the other at the same position in a context $p$ barely changes $f$'s output, the two items are treated as semantically equivalent. Formally, we calculated the semantic distance as:
 \begin{equation}
     d_s(u, v) = \mathbb{E}_{p\sim \mathcal{S}, i\sim\mathbb{U}(1,\zeta)}
     \vert F^p_i(u) - F^p_i(v) \vert
 \end{equation}
where $p$ and $i$ is a random sequence and position from the synthetic dataset $\mathcal{S}$,  respectively; $F^p_i(u)$ is the estimated probability of $u$ given the context $p[:i]$.

The \textbf{conditional distance} $d_c(u, v)$ measures how differently  $u$ and $v$ are distributed within the synthetic sequences, and can be computed directly from the synthetic data distribution $P$.
 \begin{equation}
     d_c(u, v) = \mathbb{E}_{p\sim \mathcal{S}, i\sim\mathbb{U}(1,\zeta)}\vert P_i^p(u) - P_i^p(v)\vert,
 \end{equation}
 where $P_i^p(u)$ is the conditional probability of $u$ at the $i$-th position given context $p$. The regularization loss is then defined as:
\begin{equation}
    L_{r} = \mathbb{E}_{u,v} \left( d_s(u,v) - d_c(u,v) \right) ^2,
\end{equation}
where $u$ and $v$ are uniformly sampled from $\mathcal{V}$. By minimizing $L_{r}$, we encourage the semantically similar items to have similar distribution in the synthetic dataset. 
To instantiate our method for text-based sequential recommendation, we build upon the TD3 framework~\cite{zhang2025td3} and add our regularization to the outer-loop objective $L$:
\begin{equation}
\label{eq:loss}
L(\theta, \mathcal{D}, \mathcal{S}) = L_D(\theta , \mathcal{D}) + L_{FM} (\theta, \theta_\mathcal{D}) + \lambda_{r} L_{r} (\mathcal{S})
\end{equation}
where $L$ is the total outer loop loss; $\theta_\mathcal{D}$  is the model trained on the real data $\mathcal{D}$. $L_D$ is the binary cross-entropy (BCE) loss, $L_{FM}$ is the feature matching loss as per \cite{zhang2025td3}, and $L_{r}$ and $\lambda_{r}$ is the regularization loss and balance weight.
The details about $L_D$, $L_{FM}$ and pseudo-code for the whole process are provided in Appendix \ref{sec:optimization-seqrec} and \ref{sec:pseudo-code}, respectively.



\section{Experiments}

\begin{table}[t]
    \centering
    \caption{Pre-processed dataset statistics.}
    \vspace{-0.2cm}
    \label{tab:data}
    \setlength\tabcolsep{2pt}
    \renewcommand{\arraystretch}{0.8}
    \begin{adjustbox}{max width=\linewidth}
    \begin{tabular}{lcccc}
    \toprule
    Dataset & \#Interactions & \#Items & \#Users & Avg Length \\
    \midrule
    Games & 559,128 & 35,995 & 61,639 & 9.07 \\
    Foods &  251,085 & 45,833 & 25,079 & 10.01 \\
    Yelp & 214,722 & 5,927 & 17,272 & 12.43 \\
    \bottomrule
    \end{tabular}
    \end{adjustbox}
\end{table}

\begin{table*}[th]
    \caption{Main experiment results. ``R'' and ``N'' stands for ``Recall'' and ``NDCG'', respectively. In each dataset, the best result is marked in \textbf{bold} and the second best one is \underline{underlined}. }
    \vspace{-.2cm}
    \label{tab:main}
    \centering
    \begin{adjustbox}{max width=\linewidth}
    \setlength\tabcolsep{3pt}
    \renewcommand{\arraystretch}{0.8}
    \begin{tabular}{ll|cccc|cccc|cccc}
    \toprule
    \multicolumn{2}{c|}{\multirow{2}{*}{Method}} & \multicolumn{4}{c|}{Games} & \multicolumn{4}{c|}{Foods} & \multicolumn{4}{c}{Yelp} \\
    \cmidrule(lr){3-6} \cmidrule(lr){7-10} \cmidrule(lr){11-14}
    & & R@10 & N@10 & R@20 & N@20 & R@10 & N@10 & R@20 & N@20 & R@5 & N@5 & R@10 & N@10 \\
    \midrule
    \multicolumn{2}{c|}{Full} & 0.0350 & 0.0178 & 0.0564 & 0.0232 
    & 0.0228 & 0.0108 & 0.0368 & 0.0143 
    & 0.0213 & 0.0131 & 0.0390 & 0.0188 \\
    \midrule
    \multirow{4}{48pt}{Coreset Selection} 
        & Random & 0.0317 & \underline{0.0199} & 0.0388 & 0.0217 
                 & 0.0210 & 0.0097 & 0.0295 & 0.0119
                 & 0.0133 & 0.0086 & 0.0208 & 0.0110
        \\
        & KCenter & 0.0311 & 0.0197 & 0.0380 & 0.0215 
                  & 0.0203 & 0.0095 & 0.0292 & 0.0117
                  & 0.0108 & \underline{0.0093} & 0.0133 & 0.0101
        \\
        & DEALRec & 0.0313 & \underline{0.0199} & 0.0382 & 0.0216 
                  & 0.0202 & 0.0094 & 0.0287 & 0.0115 
                  & 0.0094 & 0.0055 & 0.0152 & 0.0074
                  \\
        & GORACS & 0.0311 & 0.0195 & 0.0384 & 0.0214 
                 & 0.0204 & 0.0094 & 0.0292 & 0.0116 
                 & 0.0122 & 0.0086 & 0.0176 & 0.0103 \\
    \midrule
        \multicolumn{1}{c}{Data} & TD3 & \underline{0.0338} & 0.0185 & \underline{0.0477} & \underline{0.0219} 
        & \underline{0.0237} & \underline{0.0147} & \underline{0.0356} & \underline{0.0177}
        & \underline{0.0148} & \underline{0.0093} & \underline{0.0281} & \underline{0.0135} \\
        
        Distillation & FOSTER
        & \textbf{0.0386} & \textbf{0.0217} & \textbf{0.0524} & \textbf{0.0251} 
        & \textbf{0.0292} & \textbf{0.0148} & \textbf{0.0412} & \textbf{0.0178} 
        & \textbf{0.0210} & \textbf{0.0139} & \textbf{0.0340} & \textbf{0.0181} \\     
    \bottomrule
    \end{tabular}
    \end{adjustbox}
\end{table*}

In this section, we conduct experiments to evaluate our method. Specifically, we are interested in answering the
following research questions (RQs):
\begin{enumerate}[noitemsep, topsep=0pt, labelwidth=1.2cm, leftmargin=!]
    \item[\textbf{RQ1: }] How does our method compare against other coreset selection and dataset distillation approaches?
    \item[\textbf{RQ2: }] How does each proposed components contribute to the final performance and method efficiency?
    \item[\textbf{RQ3: }] Can our generated data be used to effectively and efficiently train other downstream backbones?
    \item[\textbf{RQ4: }] How sensitive is our method to different hyperparameters?
\end{enumerate}

Due to page limit, we defer additional experimental results to the Appendix, including: further hyperparameter sensitivity analysis on numbers of sampled items and synthetic sequences (\ref{sec:further-hyperparam-analysis}); further efficiency benchmarking for distillation and downstream training (\ref{sec:further-efficiency}); and synthetic data visualization (\ref{sec:further-visualize}). 

\subsection{Experiment Settings}

We evaluate on three datasets: Amazon Games, Amazon Foods \cite{ni2019justifying}, and Yelp~\footnote{https://www.kaggle.com/datasets/yelp-dataset/yelp-dataset} (filtered to Nashville city). All datasets are split chronologically using an absolute time-based split \cite{ji2023critical} with an 8-1-1 train-validation-test ratio. The pre-processing steps are specified in Appendix \ref{sec:data-preprocess}. For dataset distillation and coreset selection, we use 20 sequences for Amazon Games and Amazon Foods, and 60 sequences for Yelp. For all datasets, the maximum sequence length is set to 20, with sliding window pre-processing applied. Table \ref{tab:data} show the dataset statistics. The efficiency experiment is conducted on a GPU workstation with an i7-13700K CPU, NVIDIA RTX A5000 GPU, and 32GB RAM.

For the text encoder, we employ a pre-trained TinyBERT \cite{jiao2020tinybert}, and SASRec \cite{sasrec} as the recommendation backbone. 
The model is optimized using AdamW \cite{loshchilov2018decoupled} with a learning rate of 2e-4. For synthetic and coreset datasets, a weight decay of 1e-4 is applied. The synthetic sequences are optimized with a learning rate of 1e-3 and weight decay of 1e-4. The sampled item set size per step $N$ is set to 512, reset interval $R$ is set to 100, and $\lambda_r$ is set to $0.1$. Appendix \ref{sec:detailed-exp-settings} provide more experiment details for baselines and datasets.

\subsection{Main Results (RQ1)}

Table \ref{tab:main} shows the main experiment results. 
Our method achieves the best performance among all approaches and surpasses the full-dataset baseline on several metrics for Games and Foods, demonstrating that our distilled dataset can occasionally exceed full-data training -- likely due to the noise-reduction effect, a phenomenon also reported in prior work \citet{dealrec2024}. 
Furthermore, our method exhibits a pronounced advantage at smaller $K$ values: gains over the full dataset are more evident at R@10/N@10 than at R@20/N@20. 
All coreset selection methods perform similarly and fall noticeably short of the Full dataset. 
Notably, Random selection remains competitive, consistent with observations in \citet{dealrec2024, mei2025goracs}.
TD3 achieves a meaningful improvement over coreset methods and approaches full dataset performance.



\subsection{Ablation Studies (RQ2)}
\label{sec_abl}

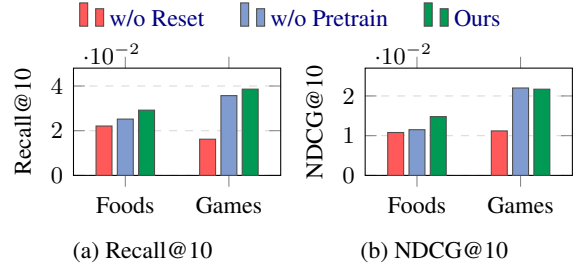
\begin{figure}[t]
    \centering
    \ref{legend_v2}
    \subcaptionbox{Recall@10}{

\begin{tikzpicture}
\begin{axis}[
    ybar,
    bar width=6pt,
    width=0.27\textwidth,
    height=3cm,
    symbolic x coords={Foods, Games},
    xtick=data,
    enlarge x limits=0.5,
    ymin=0, ymax=0.048,
    ylabel={Recall@10},
    ymajorgrids,
    grid style={dashed, black!15},
    tick label style={font=\small},
    label style={font=\small},
    legend style={
        font=\small,
        draw=none,
        fill=none,
        at={(0.5,1.02)},
        anchor=south,
        legend columns=3,
        /tikz/every even column/.append style={column sep=0.4cm},
    },
    legend to name=legend_v2
]
\addplot[fill=red!65,         draw=black!70] coordinates {(Foods,0.0221) (Games,0.0162)};
\addplot[fill=cyan!35!blue!50, draw=black!70] coordinates {(Foods,0.0252) (Games,0.0357)};
\addplot[fill=ForestGreen,          draw=black!70] coordinates {(Foods,0.0292) (Games,0.0386)};
\legend{w/o Reset, w/o Pretrain, Ours}
\end{axis}
\end{tikzpicture}}
    \hfill
    \subcaptionbox{NDCG@10}{\begin{tikzpicture}
\begin{axis}[
    ybar,
    bar width=6pt,
    width=0.27\textwidth,
    height=3cm,
    symbolic x coords={Foods, Games},
    xtick=data,
    enlarge x limits=0.5,
    ymin=0, ymax=0.027,
    ylabel={NDCG@10},
    ymajorgrids,
    grid style={dashed, black!15},
    tick label style={font=\small},
    label style={font=\small},
    legend style={
        font=\small,
        draw=none,
        fill=none,
        at={(0.5,1.02)},
        anchor=south,
        legend columns=3,
        /tikz/every even column/.append style={column sep=0.4cm},
    },
]
\addplot[fill=red!65,         draw=black!70] coordinates {(Foods,0.0108) (Games,0.0112)};
\addplot[fill=cyan!35!blue!50, draw=black!70] coordinates {(Foods,0.0115) (Games,0.0220)};
\addplot[fill=ForestGreen,          draw=black!70] coordinates {(Foods,0.0148) (Games,0.0217)};
\end{axis}
\end{tikzpicture}}
    \vspace{-0.2cm}
    \caption{Ablation study on the Foods and Games datasets. ``w/o Reset'' removes the parameter reset step; ``w/o Pretrain'' employs random initialization resets instead of pre-trained trajectory; ``Ours'' is the full method.}
    \label{fig:param-reset-results}
\end{figure}

\noindent \textbf{Reset logic. } Figure \ref{fig:param-reset-results} depicts the ablation study results for inner parameter reset logic. Without resetting the inner parameter, the performance drop drastically, demonstrating the necessity of periodically re-initializing $\theta$. Our proposed approach generally provide a better performance than only using random initialization.

\noindent \textbf{Optimization Logic. }
Table \ref{tab:rq2-optim} provides the ablation study results for various optimization approaches. 
With the first-order optimizer, when K-means (KM) is in use, instead of using on-the-fly (OTF) sampling, we select $K$ items that are closest to the centroid, and keep this selection fixed through the data distillation process. While OTF incurs a small overhead, its performance is superior as the model can see the whole item set.
Also, using all layers for data distillation demonstrates a clear performance advantage. It is worth noting that, when first-order optimizer is turned off in FOSTER, we substitute it with the bi-level BPTT approach. This bi-level approach is evaluated only under the last-layer setting as it requires higher compute resources. 
As can be seen from the results, FOSTER's efficiency enables it to scale to the all-layer setting, surpassing BPTT in both efficiency and effectiveness.


\begin{table}[]
    \centering
    \caption{Ablation on Games and Foods. We vary the bi-level optimizer (first-order approach or the bi-level BPTT), item selection (on-the-fly (OTF) or K-means (KM)), inner-problem layers (Last or All layers), and the use of regularization $L_r$. ``Time'' is the per-epoch runtime (minutes).}
    \label{tab:rq2-optim}
    \vspace{-0.2cm}
    \begin{adjustbox}{max width=\linewidth}
    \setlength\tabcolsep{2pt}
    \renewcommand{\arraystretch}{0.8}
    \begin{tabular}{cccc ccc ccc}
    \toprule
        First-
        & \multirow{2}{*}{Item} 
        & \multirow{2}{*}{Layers} 
        & \multirow{2}{*}{$L_r$}
        & \multicolumn{3}{c}{Games} 
        & \multicolumn{3}{c}{Foods} \\
        \cmidrule(lr){5-7} \cmidrule(lr){8-10}
        order& & & & R@10 & N@10 & Time & R@10 & N@10 & Time \\
    \midrule
        \xmark & OTF & Last & \checkmark & 0.0332 & 0.0181 & 9.03 
             & 0.0266 & 0.0126 & 4.77 \\
        \checkmark & OTF & Last & \checkmark & 0.0310 & 0.0202 & 1.43
             & 0.0252 & 0.0115 & 0.58 \\
        \checkmark & KM & All & \checkmark & 0.0376 & 0.0222 & 2.25
             & 0.0278 & 0.0133 & 0.98 \\
        \checkmark & OTF & All & \xmark & 0.0373 &	0.0223 & 2.33 
             & 0.0239 & 0.0114 & 0.95 \\
        \checkmark & OTF & All & \checkmark & 0.0386 & 0.0217 &  2.42
             & 0.0292 & 0.0148 & 1.02 \\
        
    \bottomrule
    \end{tabular}
    \end{adjustbox}
\end{table}

\subsection{Transfer Learning (RQ3)}

\begin{table*}[th]
    \centering
    \caption{The results of transfer learning, using TinyBERT as backbone for distilling data, while using Qwen3-4B as downstream model. ``Time'' is the total training time. ``h'' and ``m'' stands for ``hours'' and ``minutes'', respectively.}
    \begin{adjustbox}{max width=\linewidth}
    \setlength\tabcolsep{3pt}
    \renewcommand{\arraystretch}{0.8}
    \begin{tabular}{l ccc ccc ccc}
    \toprule
        Method & \multicolumn{3}{c}{Games} & \multicolumn{3}{c}{Foods} & \multicolumn{3}{c}{Yelp}  \\
        \cmidrule(lr){2-4} \cmidrule(lr){5-7} \cmidrule(lr){8-10}
        & R@10 & N@10 & Time & R@10 & N@10 & Time & R@10 & N@10 & Time \\
    \midrule
       Full  & 0.0516 & 0.0284 & 21.25h
             & 0.0456 & 0.0209 & 13.37h 
             & 0.0355 & 0.0195 & 5.77h \\
       Ours & 0.0449 & 0.0252 & 1.55h
            & 0.0435 & 0.0208 & 1.23h
            & 0.0314 & 0.0160 & 1.33h \\
        Random & 0.0229 & 0.0112 &  40.7m
               & 0.0129 & 0.0059 & 28.5m
               & 0.0207 & 0.0128 & 42.87m\\
    \bottomrule
         
    \end{tabular}
    \end{adjustbox}

    \label{tab:rq3}
\end{table*}

We conduct transfer learning experiments to evaluate the generalizability of our distilled data across different model architectures. We employ TinyBERT as the backbone for data distillation, then transfer the distilled data to train a larger downstream model, Qwen3-4B, whose detail information is specified in Appendix \ref{sec:llm-impl-details}. Table \ref{tab:rq3} shows the experiment results. Across all three datasets, our method consistently outperforms Random by a large margin on both R@10 and N@10, while remaining close to full dataset performance. This demonstrates that the distilled data retains sufficient quality to generalize well even when the downstream model differs from the one used during distillation. In terms of training efficiency, our method is considerably faster than full dataset, completing in a fraction of the time.

\vspace{-0.2cm}
\subsection{Hyperparameter Sensitive (RQ4)}
\label{sec:hyperparam_rq4}

\begin{figure}
    \centering
    \ref{lambda_r}
    \subcaptionbox{Loss weight $\lambda_r$\label{fig:rq4_lambda_r}}{
        \begin{tikzpicture}
\tikzstyle{every node}=[font=\scriptsize]
\begin{axis}[
    ybar,
    bar width=5pt,
    width=0.5\linewidth,
    height=3cm,
    enlarge x limits=0.6,
    ylabel={Recall@10},
    symbolic x coords={Foods, Games},
    xtick=data,
    ymin=0, ymax=0.055,
    yticklabel style={/pgf/number format/fixed, /pgf/number format/precision=2},
    legend style={
        font=\scriptsize,
        draw=none,
        fill=none,
        at={(0.5,1.02)},
        legend columns=3,
        /tikz/every even column/.append style={column sep=0.4cm},
    },
    legend cell align={left},
    tick align=inside,
    grid=major,
    grid style={dashed, gray!30},
    legend to name=lambda_r
]
\addplot coordinates {
    (Games, 0.0240)
    (Foods, 0.0367)
};
\addlegendentry{$\lambda_r = 0.05$}
\addplot coordinates {
    (Games, 0.0386)
    (Foods, 0.0292)
};
\addlegendentry{$\lambda_r = 0.1$}
\addplot coordinates {
    (Games, 0.0342)
    (Foods, 0.0129)
};
\addlegendentry{$\lambda_r = 0.2$}

\end{axis}
\end{tikzpicture}
        \hfill
        \begin{tikzpicture}
\tikzstyle{every node}=[font=\scriptsize]
\begin{axis}[
    ybar,
    bar width=5pt,
    width=0.5\linewidth,
    height=3cm,
    enlarge x limits=0.6,
    ylabel={NDCG@10},
    symbolic x coords={Foods, Games},
    xtick=data,
    ymin=0, ymax=0.03,
    yticklabel style={/pgf/number format/fixed, /pgf/number format/precision=2},
    tick align=inside,
    grid=major,
    grid style={dashed, gray!30},
]
\addplot coordinates {
    (Games, 0.0231)
    (Foods, 0.0131)
};
\addplot coordinates {
    (Games, 0.0217)
    (Foods, 0.0148)
};
\addplot coordinates {
    (Games, 0.0193)
    (Foods, 0.0071)
};

\end{axis}
\end{tikzpicture}
    }
    
    \ref{reset_step}
    \subcaptionbox{Reset step $R$\label{fig:rq4_R}}{
        \begin{tikzpicture}
\tikzstyle{every node}=[font=\scriptsize]
\begin{axis}[
    ybar,
    bar width=5pt,
    width=0.5\linewidth,
    height=3cm,
    enlarge x limits=0.6,
    ylabel={Recall@10},
    symbolic x coords={Foods, Games},
    xtick=data,
    ymin=0, ymax=0.055,
    yticklabel style={/pgf/number format/fixed, /pgf/number format/precision=2},
    legend style={
        font=\scriptsize,
        draw=none,
        fill=none,
        at={(0.5,1.02)},
        legend columns=3,
        /tikz/every even column/.append style={column sep=0.4cm},
    },
    legend cell align={left},
    tick align=inside,
    grid=major,
    grid style={dashed, gray!30},
    legend to name=reset_step
]
\addplot coordinates {
    (Games, 0.0249)
    (Foods, 0.0314)
};
\addlegendentry{$R = 50$}
\addplot coordinates {
    (Games, 0.0386)
    (Foods, 0.0292)
};
\addlegendentry{$R = 100$}
\addplot coordinates {
    (Games, 0.0412)
    (Foods, 0.0286)
};
\addlegendentry{$R = 150$}

\end{axis}
\end{tikzpicture}
        \hfill
        \begin{tikzpicture}
\tikzstyle{every node}=[font=\scriptsize]
\begin{axis}[
    ybar,
    bar width=5pt,
    width=0.5\linewidth,
    height=3cm,
    enlarge x limits=0.6,
    ylabel={NDCG@10},
    symbolic x coords={Foods, Games},
    xtick=data,
    ymin=0, ymax=0.03,
    yticklabel style={/pgf/number format/fixed, /pgf/number format/precision=2},
    tick align=inside,
    grid=major,
    grid style={dashed, gray!30},
]
\addplot coordinates {
    (Games, 0.0129)
    (Foods, 0.0152)
};
\addplot coordinates {
    (Games, 0.0217)
    (Foods, 0.0148)
};
\addplot coordinates {
    (Games, 0.0238)
    (Foods, 0.0140)
};

\end{axis}
\end{tikzpicture}
    }
    \vspace{-0.3cm}
    \caption{Hyperparameter sensitive analysis on Foods and Games datasets.}
    \label{fig:rq4}
\end{figure}
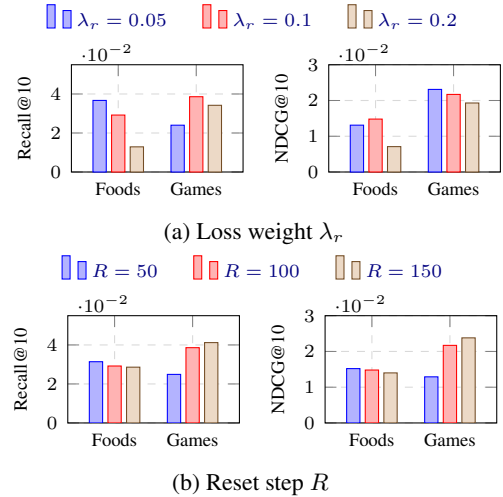

\noindent \textbf{Loss weight $\lambda_r$. } Figure \ref{fig:rq4_lambda_r} shows results on Games and Foods datasets with different $\lambda_r$. A small $\lambda_r=0.05$ caused collapse on Games, while perform well on Foods. A large $\lambda_r=0.2$ demonstrated the opposite results. The middle ground $\lambda_r=0.1$ seems to provides the best balance, suggesting it effectively regularizes without over-constraining the optimization. 

\noindent \textbf{Reset step $R$. } Figure \ref{fig:rq4_R} shows that the optimal reset step $R$ varies across datasets. On Foods, smaller values perform better, whereas on Games, larger values yield the highest Recall@10 and NDCG@10. This suggests that different datasets require different reset frequencies.

\section{Related Work}
\noindent \textbf{Text-based Sequential Recommendation. }
Text-based recommender systems leverage pretrained language models to encode rich semantic information from item textual metadata, overcoming the cold-start limitations of ID-based approaches~\cite{morec2023}.
Notable works include RecFormer~\cite{li2023text}, ONCE~\cite{liu2024once}, and others~\cite{tiger2023}, which demonstrate that modality-based recommenders are increasingly competitive with, and in some settings surpass, their ID-based counterparts.
Our work builds on this paradigm and specifically targets the training efficiency bottleneck that arises when language models are involved in the training loop.

\noindent \textbf{Efficiency for Recommender Systems. }
For training efficiency, coreset selection methods identify a representative subset of real interactions to reduce training cost, such as DEALRec~\cite{dealrec2024},  GORACS~\cite{mei2025goracs}, and NaCS~\cite{tran2026nacs}.
Dataset distillation goes further by synthesizing compact artificial training samples, with recent works such as $\infty$-AE~\cite{sachdeva2022infinite}, CGM~\cite{wang2023gradient}, TD3~\cite{zhang2025td3}, and DIET~\cite{zhang2026diet} exploring this direction in recommendation. However, all of these methods employ expensive bi-level optimization, and only TD3 focused on sequential recommendation settings.
There are other works that focuses on recommendation models training efficiency \citep{qu2025efficient}, but they are orthogonal to our works.

\noindent \textbf{Dataset Distillation.}
Dataset distillation was first introduced in~\cite{wang2018datasetdistillation} and has since been extensively studied in image classification, where approaches such as trajectory matching~\cite{cazenavette2022dataset} and distribution matching~\cite{zhao2023dataset,yin2023squeeze} have achieved strong results.
In NLP, various dataset distillation methods \citep{maekawa2024dilm, tao2024textual, shen2025condenselm} is also proposed.
However, these methods either rely on a small, fixed class space \citep{zhao2023dataset,yin2023squeeze,shen2025condenselm}, an assumption that fundamentally breaks down in recommendation, or rely on an expensive bi-level optimization \citep{maekawa2024dilm,cazenavette2022dataset,zhang2025td3}.
For further reading, we refer the reader to~\cite{li2022awesome}.

\section{Conclusion}
In this paper, we propose FOSTER, namely first-order dataset distillation for text-based sequential recommendation. 
Our proposed approach includes three main components to further improve the efficiency and the performance. First, instead of using full item set for each timestep, we employ an on-the-fly approach, where we will randomly sample a smaller subset of items. Second, we employ BOME with periodical re-anchor to the inner parameters by the pre-train trajectory checkpoints, drastically reducing the compute resource requirement, while ensuring the synthetic dataset performance. Third, we introduce a regularization loss based on distributional hypothesis to further enhance the final results.
Extensive experiments show that our method is not only more efficient but also more accurate than previous state-of-the-art approaches.

\section*{Limitations}
While our method demonstrates strong empirical results across multiple benchmarks, we identify two limitations that warrant further investigation.

\noindent \textbf{Hyperparameter sensitivity.} Our approach introduces several hyperparameters, including the number of inner steps $T$, the reset interval $R$, and the number of items sampled each step $N$. Although we provide sensitivity analyses in Sec. \ref{sec:hyperparam_rq4}, selecting these values in practice requires validation effort. Developing more principled or adaptive strategies for hyperparameter selection remains an open direction.

\noindent \textbf{Performance gap and limited LLM transferability}. There is still a small performance gap between full dataset and the synthetic dataset in Yelp dataset. Furthermore, while our distilled data transfers effectively to conventional sequential models, the gap between the full dataset and the synthetic dataset for a large language model on the Games and Yelp datasets are still large. We hypothesize this is partly due to the representational gap between the source backbone and the target LLM models, and regard closing this gap as a promising avenue for future work.


\bibliography{custom}

\appendix
\label{sec:appendix}

\section{Theoretical Details}
\label{app:first_order}


Let $L_{\mathrm{in}}(\mathcal{S},\theta)$ denote the inner loss on the synthetic dataset
$S$ with model parameter $\theta$. The inner condition
$\theta=\theta^*(\mathcal{S}) \in \arg\min_{\theta'} L_{\mathrm{in}}(\mathcal{S},\theta')$ enforces $\theta$
to be optimal on $S$. Let
\begin{equation}
    L_{\mathrm{in}}^*(\mathcal{S})
    =
    \min_{\theta'} L_{\mathrm{in}}(\mathcal{S},\theta')
    \label{eq:inner_value}
\end{equation}
be the minimum achievable inner loss. Since
$L_{\mathrm{in}}(\mathcal{S},\theta) \geq L_{\mathrm{in}}^*(\mathcal{S})$ for any $\theta$, the
inner optimality condition can be written as the constraint
\begin{equation}
\label{eq:bome-obj2}
\min_{\mathcal{S}, \theta}\;
L_{\mathrm{out}}(\theta, \mathcal{S}, \mathcal{D})
\quad \text{s.t.} \quad q(\mathcal{S}, \theta) \leq 0,
\end{equation}
Because $q(\mathcal{S},\theta)$ is always non-negative, this constraint is satisfied only
when $\theta$ reaches the inner optimum on $S$.

To optimize Eq.~\eqref{eq:bome-obj2}, we need the constraint gradient
\begin{equation}
    \nabla_S q(\mathcal{S},\theta)
    =
    \nabla_1 L_{\mathrm{in}}(\mathcal{S},\theta)
    -
    \nabla_S L_{\mathrm{in}}^*(\mathcal{S}),
    \label{eq:q_gradient}
\end{equation}
where the second term depends on the inner optimum $\theta^*(\mathcal{S})$. Directly
computing this term would seem to require the implicit gradient
$\nabla_S\theta^*(\mathcal{S})$. 
Next, we will show that this dependence vanishes under the inner first-order optimality condition. By the chain rule,
\begin{equation}
\begin{aligned}
    \nabla_S L_{\mathrm{in}}^*(\mathcal{S})
    &=
    \nabla_1 L_{\mathrm{in}}(\mathcal{S},\theta^*(\mathcal{S})) \\
    &+
    J_{\theta^*}(\mathcal{S})^\top
    \nabla_2 L_{\mathrm{in}}(\mathcal{S},\theta^*(\mathcal{S})),
\end{aligned}
\end{equation}
where $J_{\theta^*}(\mathcal{S})$ is the Jacobian of $\theta^*(\mathcal{S})$ with respect to $S$.
Since $\theta^*(\mathcal{S})$ is an inner optimum, the first-order optimality condition
gives
\[
    \nabla_2 L_{\mathrm{in}}(\mathcal{S},\theta^*(\mathcal{S}))=0.
\]
Therefore, the second term vanishes, yielding
\[
    \nabla_S L_{\mathrm{in}}^*(\mathcal{S})
    =
    \nabla_1 L_{\mathrm{in}}(\mathcal{S},\theta^*(\mathcal{S})).
\]
Thus, the value-function gradient does not require differentiating through
$\theta^*(\mathcal{S})$, which is also a direct consequence of Danskin's theorem \cite{danskin2012theory}.
Therefore, Eq. \ref{eq:bome-obj2} can be solved by first-order gradients alone.
\vspace{-0.2cm}
\section{Sequential Recommendation Dataset Distillation Optimization Objective}
\label{sec:optimization-seqrec}
The distillation is framed as a bi-level optimization problem where the inner loop trains the sequential encoder on the synthetic summary and the outer loop evaluates it on the real data. Concretely, the inner objective employs KL divergence between the model's predicted distribution and the synthetic soft targets, while the outer loop will optimize the actual objective -- improving the synthetic data $\mathbf{S}$. In this section, we will introduce about the optimization process of outer and inner loops.

\subsection{Outer Loop}
Following standard practice in sequential recommendation \cite{sasrec}, we employ a binary cross-entropy formulation $L_{\mathcal{D}}$ over autoregressive next-item predictions as the recommendation model's training objective, given by:
\begin{equation}
\begin{aligned}
    L_{\mathcal{D}} &= \sum_{i=1}^N \big( BCE(\langle \hat{\mathbf{e}}_i, \mathbf{e}_i \rangle, 1) \\ 
    &+ BCE(\langle \hat{\mathbf{e}}_i, \mathbf{e}^{neg}_i \rangle, 0) \big),
\end{aligned}
\end{equation}
where $\hat{\mathbf{e}}_i$ is the contextual representation produced at position $i$ used to predict the next item, $\langle \cdot ,\cdot \rangle$ is dot product, $\mathbf{e}^{neg}_i$ is the negatively sampled samples at i-th timestep.

\citet{zhang2025td3} extends the outer loop objective beyond naive performance matching by incorporating a feature space alignment term:
\begin{equation}
\label{{eq:opt-obj-1}}
    \begin{aligned}
     \mathbf{S}^* &= \arg\min_{\mathbf{S}} \; \mathcal{L}_{\mathcal{D}}(\theta^*,  \mathcal{D}) + L_{FM} \\
     L_{FM} &= \frac{1}{2} \left\| f(\theta_\mathcal{D}, \mathcal{D}) - f(\theta^*, \mathcal{D}) \right\|_2^2, \\
    \text{ s.t. } \theta^* &= \arg\min_\theta \mathcal{L}_{\text{in}}(\theta, \mathbf{S})
    \end{aligned}
\end{equation}

where $L_{FM}$ is feature matching loss, $f(\theta_\mathcal{D}, \cdot)$ and $f(\theta^{*}, \cdot)$ are the recommendation models trained on the real and synthetic data, respectively. This alignment loss encourages the synthetic-data-trained model to converge to a similar feature space solution as the real-data-trained model, promoting a more favorable loss landscape. The bilevel problem is solved via RaT-BPTT. 

\subsection{Inner Loop}

In the inner loop, we first define:
\begin{equation}
\begin{aligned}
    \mathbf{X}_i &= \mathbf{S}[i, :\zeta, :], \mathbf{y}_i = \mathbf{S}[i, \zeta], 
\end{aligned}
\end{equation}
where $\mathbf{S}[i, :, :\zeta, :]$ follows NumPy slice notation, $\mathbf{X}_i \in \mathbb{R}^{\left( \zeta  - 1 \right) \times \vert \mathcal{V} \vert}$ is the input, $\mathbf{y}_i \in \mathbb{R}^{\vert \mathcal{V} \vert}$ denotes the synthetic target distribution. 
As $\mathbf{X}_i$ represents a probability distribution over items rather than a discrete item sequence, it cannot be directly processed by the recommendation model $f$. We therefore compute the embedding of $\mathbf{X}_i$ as a weighted average of all (in TD3 \cite{zhang2025td3}) or sampled (in FOSTER) item embeddings $\mathbf{E}$, with weights given by $\mathbf{X}_i$:
\begin{equation}
    \mathbf{E}'_i = l(\mathbf{X}_i) = \langle \mathbf{X}_i, \mathbf{E} \rangle,
\end{equation}
where $l$ is the language model, $\mathbf{E}$ is the embedding of all (sampled) items. Thus, we can process $\mathbf{E}'_i$ normally by the recommendation model $g$, producing output representation $\mathbf{h}_i = g(\mathbf{E}'_i) = f(\mathbf{X}_i)$. Then, the final output:
\begin{equation}
    \mathbf{\hat{y}} = \mathbf{h}^\top_i \mathbf{E}'_i
\end{equation}
Finally, we employ KL Divergence instead of BCE as the loss function for the inner training loop:
\begin{equation}
\label{eq:inner-obj}
    L_{in} (\theta, \mathcal{S}) = D_{KL} (\mathbf{y}_i \Vert \mathbf{\hat{y}_i}).
\end{equation}
Using KL divergence is natural in this setting because each synthetic position is not assigned to a single discrete item; instead, it represents a soft supervisory signal that may place non-trivial probability mass on multiple semantically related items.

Algorithm \ref{alg:train-synth} shows this process.
We first sampled a smaller item set $\mathcal{V}_k$ (line \textbf{1}). Then, we get the embedding $\mathbf{E}_k$ from $\mathcal{V}_k$ (line \textbf{2}), note that $\mathbf{E}_k$ is a pre-computed embedding table using a pre-trained model, not the embedding table extracted from the currently in-training $l$. Then, we sample the input $\mathbf{X}_i$ and the target $\mathbf{y}_i$ (Line \textbf{3}). Then, we perform standard pipeline of forward (line \textbf{4-5}), calculating loss (line \textbf{6}) and update model (line \textbf{7}).

\begin{algorithm}[th]
\caption{A training step on the synthetic dataset}
\label{alg:train-synth}
\KwData{Synthetic data tensors $\mathbf{G}, \mathbf{U}, \mathbf{T}$, embedding table $\mathbf{E}$, learning rate $\alpha$, Real item list $\mathcal{V}$}
\SetKwComment{Comment}{\# }{}


    $\mathcal{V}_k \gets $ Random sample from $\mathcal{V}$\;
    $\mathbf{E}'_k \gets $ Get embedding corresponding for $\mathcal{V}_k$\;
    $\mathbf{X}_i, \mathbf{y}_i \gets \llbracket \mathbf{G}, \mathbf{U}[i], \mathbf{T}, \mathbf{E}_k \rrbracket$\;
    $\mathbf{E}'_k \gets l(\mathcal{V}_k)$\;
    $\mathbf{\hat{y}}_k \gets g(\mathbf{S} \odot \mathbf{E}'_k)$\; 
    $L \gets $ Eq. \ref{eq:inner-obj}\;
    $g,l \gets g - \alpha\nabla_g L, l - \alpha\nabla_lL$\;


\end{algorithm}

\section{Pseudo-code}
\label{sec:pseudo-code}

\begin{algorithm}[th]
\caption{The proposed dataset distillation algorithm (FOSTER)}
\label{alg:main}
\KwData{Real data $\mathcal{D}$, learning rate $\alpha$, inner parameter reset intervals $R$, number of inner loop steps $T$}
\SetKwComment{Comment}{\# }{}


\Comment{\color{MidnightBlue} Prepare step}
Train $\theta$ on $\mathcal{D}$ to get pre-train trajectory\;

$\mathbf{E} \gets $ Encode all real items $\mathcal{V}_\mathcal{D}$\;

\Comment{\color{MidnightBlue} Main dataset distillation step}
Initialize $\mathcal{S}$ ($\{\mathbf{T}, \mathbf{G}, \mathbf{U}\}$)\;
Iteration $i \gets 0$\;
\Repeat{converged}{
    \If{$i \mod R = 0$}{
        $\theta \gets$ Sample from the pre-train trajectory\;
    }
    \Comment{First-order update (Sec. \ref{sec:bome})}
    $\theta^* \gets $ Train $\theta$ on $\mathcal{S}$ for T steps\;

    $q(\mathcal{S}, \theta) \gets L_S(\theta, \mathcal{S}) - L_S(\theta^*, \mathcal{S})$\;
    $\lambda_k \gets $ Eq. \ref{eq:lambdk}\;
    $L \gets$ Eq. \ref{eq:loss}\;
    $(\mathcal{S}, \theta) \gets (\mathcal{S}, \theta) - \alpha (\nabla L + \lambda_k \nabla q )$\;
    
    \Comment{End first-order update}
    $i \gets i + 1$\;
}
\end{algorithm}

Algorithm \ref{alg:main} shows the pseudo-code for the proposed algorithm. First, we need to train the model on the real dataset (Line \textbf{1}). This prepare step is common in various dataset distillation approaches, such as MTT \cite{cazenavette2022dataset} and TD3 \cite{zhang2025td3}.  Then, we encode item embedding (Line \textbf{2}) to get the item embedding used in calculating $\mathbf{S}$. Then, we can start to distill the synthetic dataset. 
At each iteration, we first reset $\theta$ every $R$ steps from pre-trained trajectory; (Line \textbf{6-7}).
Then, we train $\theta$ on $\mathcal{S}$ for T steps to estimate $\theta_{\mathcal{S}}$ (Line \textbf{8}), detailed in Alg. \ref{alg:train-synth}. Note that in these T steps, we will sample a different item set for each step. After that, we compute constraint $q(\mathcal{S}, \theta)$ (Line \textbf{9}); and update $(\mathcal{S}, \theta)$ via the first-order gradient step (Line \textbf{10-12}). The process is iterated until convergence. 

\section{Detailed Experiment Settings}
\label{sec:detailed-exp-settings}

\subsection{Baselines}

In this section, we provide details on baselines used in the main experiments.
\begin{itemize}
    \item \textbf{Random} randomly samples a subset of users and uses their interaction as the coreset.
    \item \textbf{K-Center} represents each user by their last-timestep embedding and applies the greedy $k$-center algorithm to select the subset.
    \item \textbf{DEALRec}~\cite{dealrec2024} scores each training sample using an effort score derived from a pre-trained model, pruning low-effort and redundant sequences.
    \item \textbf{GORACS}~\cite{mei2025goracs} selects coresets by bounding the test loss via optimal transport distance and gradient information, avoiding repeated model retraining during selection.
    \item \textbf{TD3}~\cite{zhang2025td3} generates a compact set of pseudo-sequences via Tucker decomposition to match the distributional statistics of the full training set. For TD3, due to the high computational cost, we only unfreeze the last layer during the inner loop training. 
\end{itemize}

\subsection{Data}
\label{sec:data-preprocess}

We source all datasets from RecBole \cite{recbole121} and apply standard pre-processing steps. Interaction records are first sorted chronologically by timestamp, then filtered using the 5-core setting (retaining only users and items with at least 5 interactions). For Yelp, we additionally restrict items to those located in Nashville city prior to applying the 5-core filter. During training, we apply a sliding window strategy to generate additional samples from long interaction sequences \cite{mei2025goracs}, while at evaluation time each sequence is truncated to the last 20 interactions. 

\subsection{Implementation Details}
The inner steps $T$ is searched in $\{40, 50, 60\}$. The model and synthetic data is selected with the best results in validation dataset in R@10. We calculate metrics using all items.

\subsection{LLM Implementation Details}
\label{sec:llm-impl-details}

Follow \cite{liu2024once}, we employ LLM (Qwen3-4B \cite{yang2025qwen3}) directly as a text encoder. We apply LoRA with rank 32 and alpha 128, yielding a scaling factor of 4 to the last two transformer layers (layers 34 and 35), with dropout of 0.1. All of the other layers are frozen. To speed up the training process, we adopt a caching strategy wherein we pre-compute and store the hidden states from the first layers for all item embeddings. 

\section{Additional Experiment Results}
\label{sec:further-exp-res}

\subsection{Further Hyperparameter Analysis}
\label{sec:further-hyperparam-analysis}
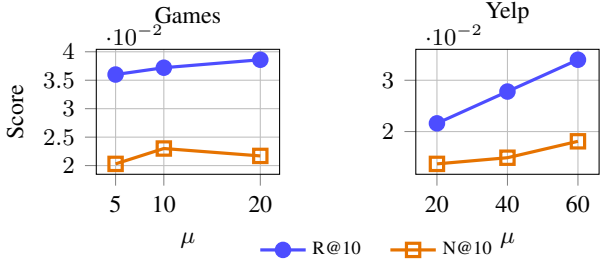
\begin{figure}
    \centering
    \begin{tikzpicture}
\begin{groupplot}[
    group style={
        group size=2 by 1,
        horizontal sep=1.8cm,
    },
    width=0.52\columnwidth,
    height=0.42\columnwidth,
    grid=both,
    grid style={line width=0.3pt, draw=gray!30},
    major grid style={line width=0.4pt, draw=gray!50},
    tick align=outside,
    tick pos=left,
    label style={font=\small},
    tick label style={font=\small},
    legend style={
        font=\scriptsize,
        at={(0.5,1.02)},
        anchor=south,
        legend columns=2,
        /tikz/every even column/.append style={column sep=0.3cm},
        draw=none,
        fill=none,
    },
    xlabel={$\mu$},
]

\nextgroupplot[
    xtick={5, 10, 20},
    xticklabels={5, 10, 20},
    xmin=3, xmax=22,
    ylabel={Score},
    legend to name=sharedlegend,
    title={\small Games},
    legend entries={R@10, N@10, Full R@10, Full N@10},
]

\addplot[
    color=blue!70,
    mark=*,
    mark size=2.5pt,
    line width=1.2pt,
] coordinates {
    (5,  0.0360)
    (10, 0.0372)
    (20, 0.0386)
};

\addplot[
    color=orange!90!black,
    mark=square,
    mark size=2.5pt,
    line width=1.2pt,
] coordinates {
    (5,  0.0203)
    (10, 0.0230)
    (20, 0.0217)
};

\nextgroupplot[
    xtick={20, 40, 60},
    xticklabels={20, 40, 60},
    xmin=14, xmax=66,
    title={\small Yelp},
]

\addplot[
    color=blue!70,
    mark=*,
    mark size=2.5pt,
    line width=1.2pt,
] coordinates {
    (20, 0.0216)
    (40, 0.0278)
    (60, 0.0340)
};

\addplot[
    color=orange!90!black,
    mark=square,
    mark size=2.5pt,
    line width=1.2pt,
] coordinates {
    (20, 0.0137)
    (40, 0.0149)
    (60, 0.0181)
};

\end{groupplot}

\node at ($(group c1r1.south)!0.5!(group c2r1.south) + (0.4,-1.0)$)
    {\pgfplotslegendfromname{sharedlegend}};

\end{tikzpicture}
    \vspace{-0.8cm}
    \caption{Model performance under varying numbers of distilled sequences $\mu$ on Games and Yelp datasets.}
    \label{fig:num_seq}
\end{figure}

\noindent \textbf{Number of sequences. } Figure \ref{fig:num_seq} shows the experiment results on how the number of sequences affects the downstream model performance.  

\begin{figure}
    \centering
    \ref{item_k}
    {
        \begin{tikzpicture}
\tikzstyle{every node}=[font=\scriptsize]
\begin{axis}[
    ybar,
    bar width=5pt,
    width=0.5\linewidth,
    height=3cm,
    enlarge x limits=0.6,
    ylabel={Recall@10},
    symbolic x coords={Foods, Games},
    xtick=data,
    ymin=0, ymax=0.055,
    yticklabel style={/pgf/number format/fixed, /pgf/number format/precision=2},
    legend style={
        font=\scriptsize,
        draw=none,
        fill=none,
        at={(0.5,1.02)},
        legend columns=3,
        /tikz/every even column/.append style={column sep=0.4cm},
    },
    legend cell align={left},
    tick align=inside,
    grid=major,
    grid style={dashed, gray!30},
    legend to name=item_k
]
\addplot coordinates {
    (Games, 0.0388)
    (Foods, 0.0194)
};
\addlegendentry{$N = 256$}
\addplot coordinates {
    (Games, 0.0386)
    (Foods, 0.0292)
};
\addlegendentry{$N = 512$}
\addplot coordinates {
    (Games, 0.0312)
    (Foods, 0.0334)
};
\addlegendentry{$N=1024$}

\end{axis}
\end{tikzpicture}
        \hfill
        \begin{tikzpicture}
\tikzstyle{every node}=[font=\scriptsize]
\begin{axis}[
    ybar,
    bar width=5pt,
    width=0.5\linewidth,
    height=3cm,
    enlarge x limits=0.6,
    ylabel={NDCG@10},
    symbolic x coords={Foods, Games},
    xtick=data,
    ymin=0, ymax=0.03,
    yticklabel style={/pgf/number format/fixed, /pgf/number format/precision=2},
    tick align=inside,
    grid=major,
    grid style={dashed, gray!30},
]
\addplot coordinates {
    (Games,  0.0236)
    (Foods, 0.0095)
};
\addplot coordinates {
    (Games, 0.0217)
    (Foods, 0.0148)
};
\addplot coordinates {
    (Games, 0.0175)
    (Foods, 0.0158)
};

\end{axis}
\end{tikzpicture}
    }
    \caption{The effect of number of item sampled per step $K$ on the performance.}
    \label{fig:item_k}
\end{figure}
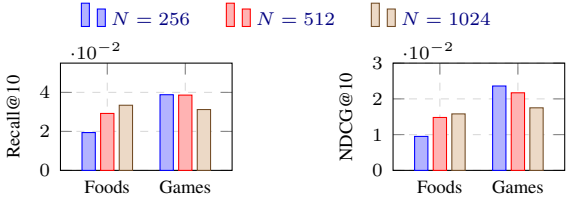

\noindent \textbf{Number of sampled items. }  Figure \ref{fig:item_k} shows the experiment results on how the number of sampled items per step $N$ affects the downstream model performance.  

\subsection{Efficiency Benchmark}
\label{sec:further-efficiency}

\begin{table}[ht]
\centering
\caption{Runtime (minutes) and peak GPU memory usage (MB) per epoch on the Foods dataset, measured on a single NVIDIA RTX A5000 (24GB VRAM). OOM indicates out-of-memory.}
\vspace{-0.2cm}
\label{tab:efficiency_foods}
\begin{adjustbox}{max width=\linewidth}
\setlength{\tabcolsep}{3pt}
\renewcommand{\arraystretch}{0.8}
\begin{tabular}{llcrr}
\toprule
Method & \#Items & Layers & Time & Mem \\
\midrule
TD3 & All   & Last & OOM   & 33,002 \\
FOSTER & All   & Last & 10.43 & 20,654 \\
TD3 & 512   & Last & 2.20  &  5,004 \\
FOSTER & 512   & Last & 0.58  &  1,296 \\
TD3 & 512   & All  & 5.07  & 19,368 \\
FOSTER & 512   & All  & 0.98  &  5,198 \\
\bottomrule
\end{tabular}
\end{adjustbox}
\end{table}

Table~\ref{tab:efficiency_foods} reports per-epoch runtime and peak GPU memory on Foods. TD3 \cite{zhang2025td3} immediately exceeds the 24GB VRAM budget, confirming that naive bi-level optimization does not scale to large catalogues. Reducing to a 512-item subset via stochastic sampling yields the largest gains: FOSTER requires only 1,296MB and 0.58 minutes per epoch --- a $25.46\times$ memory reduction over TD3. Even under all-layer fine-tuning, FOSTER (5,198MB, 0.98 minutes) remains substantially cheaper than TD3 (19,368MB, 5.07 minutes), demonstrating that first-order optimization is critical for tractable training under consumer-grade GPU constraints.


\begin{table}[t]
\caption{Training time comparison between full training and the synthetic data on Foods dataset. ``h'' and ``m'' stands for hours and minutes, respectively.}
\label{tab:efficiency-foods}
\centering
\begin{adjustbox}{max width=\linewidth}
\setlength\tabcolsep{3pt}
\renewcommand{\arraystretch}{0.8}
\begin{tabular}{l|ccc|ccc}
\toprule
Backbone         & \multicolumn{3}{c|}{Full}      & \multicolumn{3}{c}{Ours} \\
    \cmidrule(lr){2-4} \cmidrule(lr){5-7}
         & R@10   & N@10   & Time        & R@10   & N@10   & Time       \\
\midrule
TinyBERT & 0.0228 & 0.0108 & 22.2m     & 0.0292 & 0.0148 & 3.7m     \\
Qwen3-0.6B & 0.0292 & 0.0148 & 7.5h   & 0.0368 & 0.0179 & 54.2m    \\
Qwen3-4B  & 0.0391 & 0.0182 & 13.4h & 0.0435 & 0.0208 & 1.2h \\
\bottomrule
\end{tabular}
\end{adjustbox}
\end{table}

Table \ref{tab:efficiency-foods} shows the training efficiency on Foods dataset. Our method successfully not only approximate the performance of the full training, but also drastically reduce the required training time.

\subsection{Further results analysis}

\begin{figure}
    \centering
    \includegraphics[width=\linewidth]{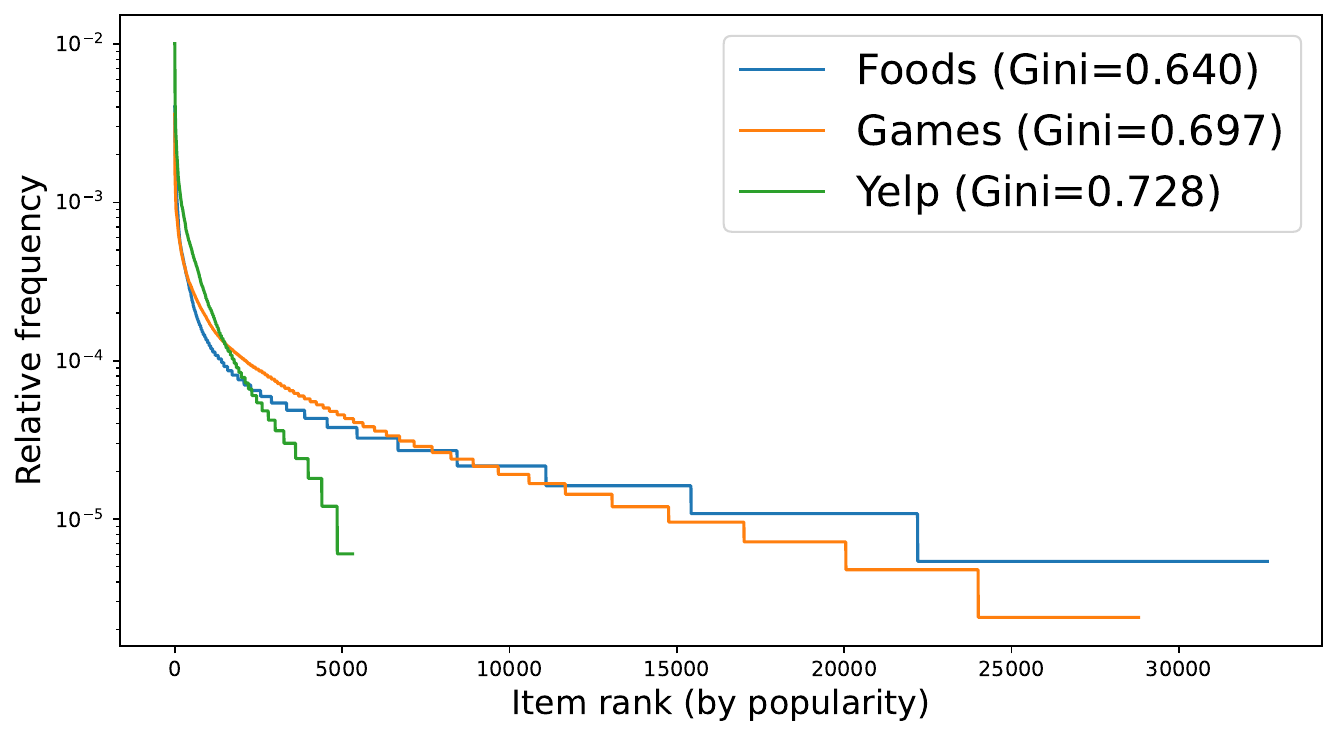}
    \caption{Item popularity distributions of the three training datasets.}
    \label{fig:popularity}
\end{figure}

In this section, we further analyze the lower results on Yelp. Figure \ref{fig:popularity} illustrates the normalized item popularity distributions across three datasets. All three datasets exhibit a power-law decay, where a small fraction of items account for the majority of interactions. However, the datasets differ substantially in catalog size and tail behavior. Both Foods and Games maintain extended long tails spanning over 30,000 items. In contrast, Yelp has a substantially smaller catalog (~5,000 items) with a steeper decay and an abrupt tail cutoff. Because Yelp's naturally short and sparse tail provides almost no gradient signal for tail items during optimization process, the synthetic sequences collapse toward the popular items.

\subsection{Data Visualization}
\label{sec:further-visualize}

We calculate the probabilities of all items, and for each user, we visualized the top 10 items that has highest probabilities. 

\begin{figure*}[h]
    \centering
    \includegraphics[width=0.9\linewidth]{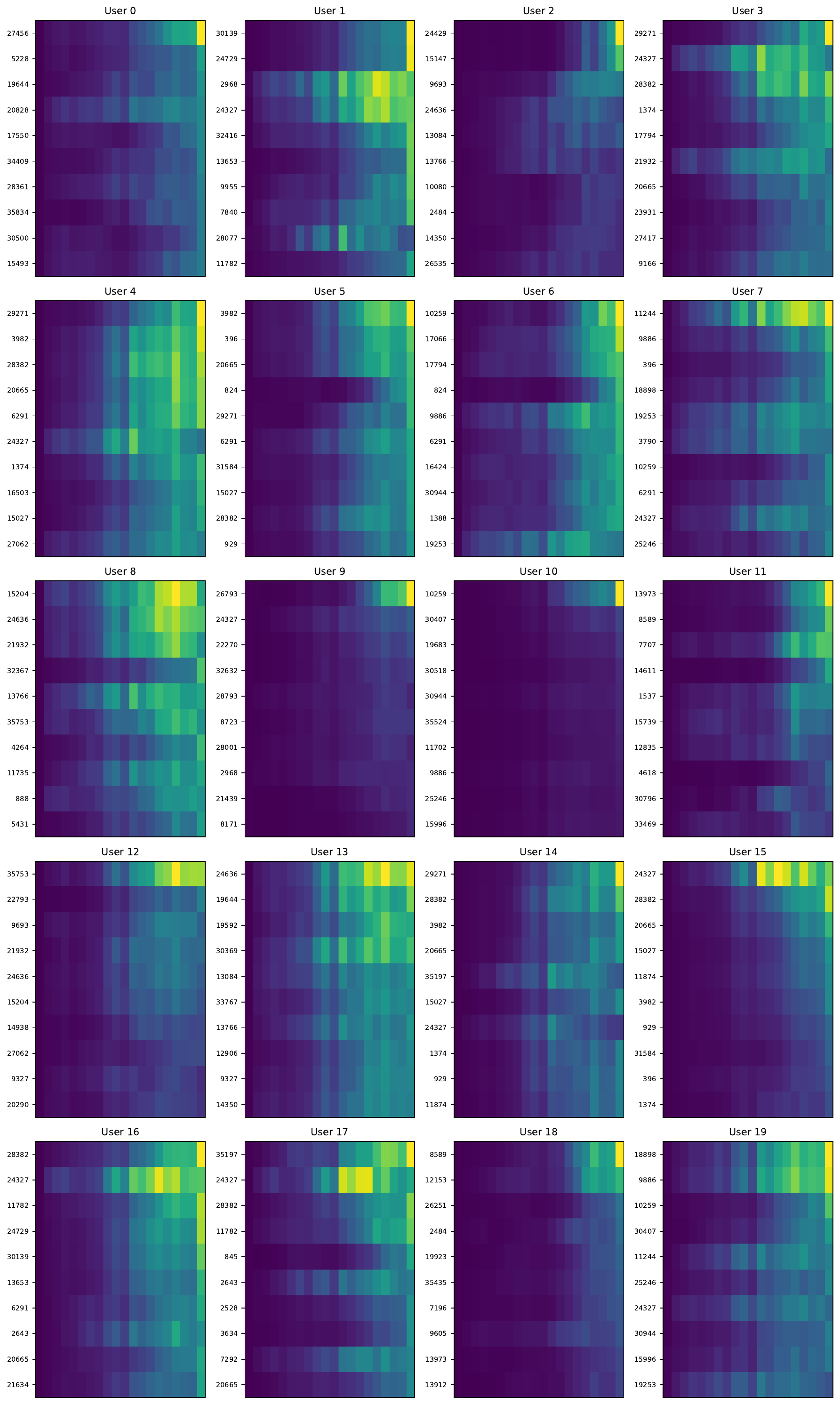}
    \caption{Games dataset visualization}
\end{figure*}

\begin{figure*}[h]
    \centering
    \includegraphics[width=0.9\linewidth]{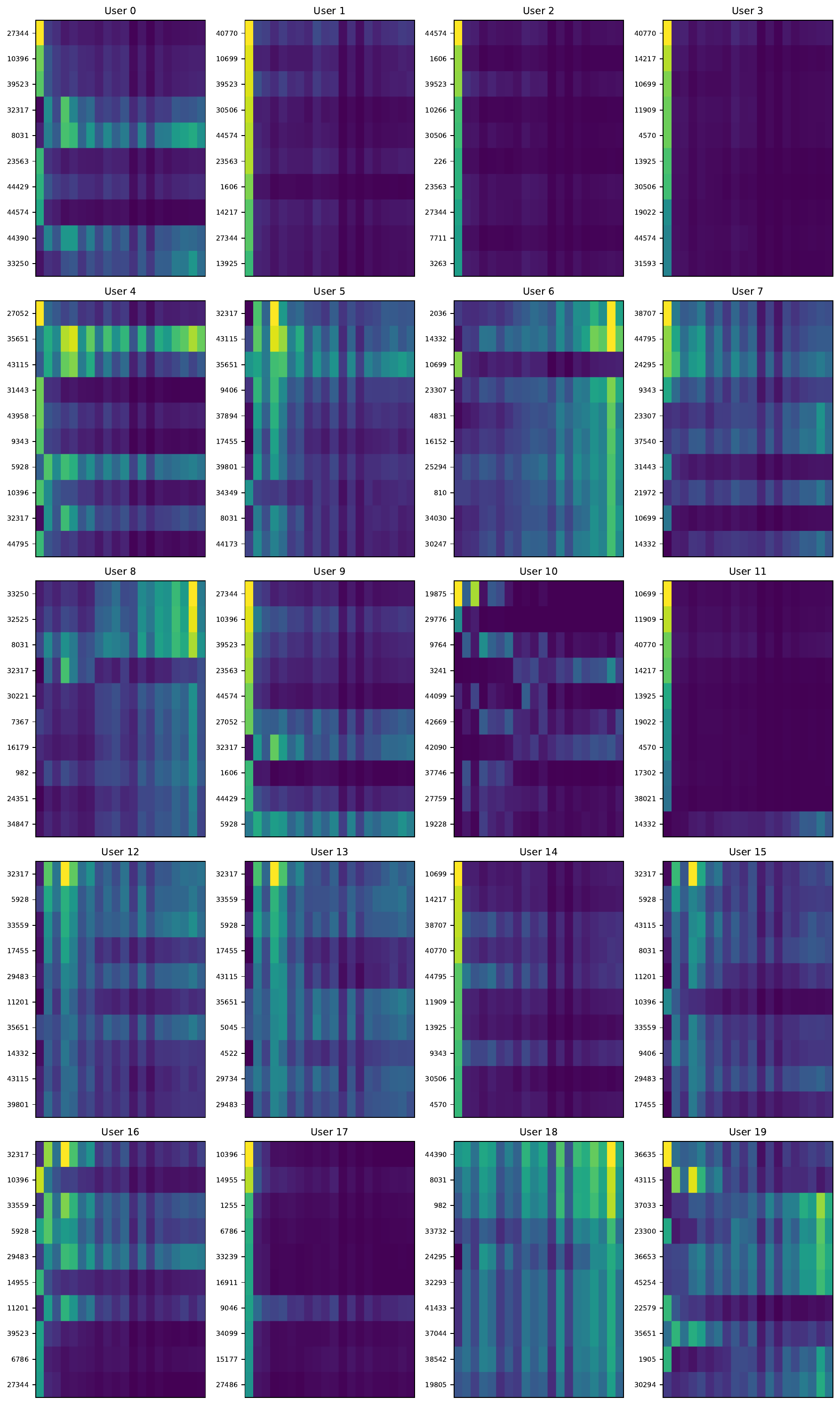}
    \caption{Foods dataset visualization}
\end{figure*}



\end{document}